\documentclass[12pt]{article}
\usepackage{epsfig,amsmath,amssymb,graphics}

\newcommand{\LL}{\mathcal{L}}
\newcommand{\Tr}{\text{ Tr }}

\newcommand{\lsim}{\,\raise.3ex\hbox{$<$\kern-.75em\lower1ex\hbox{$\sim$}}\,}
\newcommand{\gsim}{\,\raise.3ex\hbox{$>$\kern-.75em\lower1ex\hbox{$\sim$}}\,}
\begin{document}

\begin{titlepage}
\begin{flushright}
HUTP-02/A031\\
hep-ph/0207164\\
\end{flushright}
\vskip 2cm
\begin{center}
{\large\bf Deconstructing six dimensional gauge theories with strongly
coupled moose meshes  
}
\vskip 1cm
{\normalsize
\mbox{
\hspace{-0.3in}
Thomas Gregoire and Jay G. Wacker}\\
\vskip 0.5cm

Department of Physics, University of California\\
Berkeley, CA~~94720, USA\\
and \\
Theory Group, Lawrence Berkeley National Laboratory\\
Berkeley, CA~~94720, USA\\
}
\vspace{0.2in}
Jefferson Physical Laboratory\\
Harvard University\\
Cambridge, MA 02138\\
\vskip .3cm

\end{center}

\vskip .5cm

\begin{abstract}
It has recently been realized that five dimensional theories can be
generated dynamically from asymptotically free, QCD-like four dimensional
dynamics via ``deconstruction.''  In this paper we generalize this
construction to six dimensional theories using a moose mesh with
alternating weak and strong gauge groups.  A new ingredient is the
appearance  of self couplings  between the higher dimensional components of
the gauge fields that appear as a potential for pseudo-Goldstone bosons in
the deconstructed picture. We show that, in the limit where the weak gauge
couplings are made large, such potentials are generated with
appropriate size from finite one loop correction.  Our construction has a number of
applications, in particular to the constructions of ``little Higgs'' models of electroweak symmetry breaking.
\end{abstract}

\end{titlepage}

\section{Introduction}
Field theories in higher dimensions are useful to address various issues in particle
physics (for some examples see
\cite{Arkani-Hamed:1998rs,Antoniadis:1998ig,Arkani-Hamed:1998nn,Antoniadis:1990ew,Randall:1999ee,Randall:1999vf,Arkani-Hamed:1999yy,Arkani-Hamed:1998sj,Arkani-Hamed:1999pv,Arkani-Hamed:1999dc,Barbieri:2000vh,Hall:2001pg}).
These higher dimensional field theories have dimensionful coupling constants
much like gravity in four dimensions and become strongly coupled at
high energies requiring a new description of the physics.
In \cite{Arkani-Hamed:2001ca} it was shown that five dimensional Yang-Mills 
theories compactified on a circle could be generated dynamically from  asymptotically free, four dimensional theories by
``deconstructing'' the extra dimension. We first latticize the circle
\cite{Hill:2000mu,Cheng:2001vd}, and the latticized action then take the form
of a gauged non-linear sigma model. 
The gauged non-linear sigma model breaks the large, product gauge 
symmetry down to the diagonal subgroup and can be represented 
graphically as theory space (or ``moose'' or quiver diagram).
In the continuum limit theory space becomes the actual extra dimension.
The theory space structure is very rich and has many applications \cite{Arkani-Hamed:2001nc,Arkani-Hamed:2001vr,Arkani-Hamed:2001ed,Arkani-Hamed:2001ie,Cheng:2001an,Skiba:2002nx}

The non-linear sigma model is non-renormalizable and breaks down at high
energy, therefore  simply latticizing the extra dimension does not provide an
ultraviolet description of five dimensional theories. 
However, the non-linear sigma
model can be easily UV completed in standard ways, for example with a
strongly coupled theory, \`a la QCD.
 At very high
energy, the theory is purely four dimensional and asymptotically free. At
some scale $\Lambda$, the strongly interacting gauge group condenses, breaking
chiral symmetries, and at low energies the 
description is in terms of non-linear sigma model fields that looks
exactly like a five dimensional theory compactified on a latticized circle. 
Beneath the scale of the lattice spacing and above the
compactification radius the theory looks five dimensional.

Generating a six
dimensional theory compactified on a torus in a similar way is not as
straight forward. The first step is to latticize the torus in a 2
dimensional grid of points and links, where the points are gauge groups and
the links are bifundamental non linear sigma model fields. The non linear
sigma model fields arise from fermion condensates triggered by  strong gauge
groups.  Kinetic terms for these  fields correspond to $\Tr F_{\mu 5}^2$
and $\Tr F_{\mu 6}^2$ terms of the continuum theory. The issue is then to
generate  the $\Tr F_{56}^2$ operator which would  appear
as a potential for the non-linear sigma model fields in the latticized theory.  Naively this appears
to be at least an eight fermion operator in the ultraviolet theory and the
coefficient of such term appears to be too small \cite{Lane:2002pe}. 
We demonstrate in this paper that these terms are in fact generated with
sizeable coefficients from the low energy theory in the limit where the
``weak'' gauge couplings are made large. The scaling of couplings and the structure
of radiative corrections  to the effective action are crucial for
understanding how this arise. We will see that the one loop finite
contributions from gauge interactions in the deconstructed theory  will be
sufficient to generate sizeable potentials for the non-linear sigma model fields.

In section \ref{Sec: UV Completion}, we review the principles of
dimensional deconstruction. In section \ref{Sec: Scaling}, we present the relations between the
parameters of the extra-dimensional theory we wish to generate and the
 parameters of the four dimensional non-linear sigma model from which it emerge. We also define the
limit we wish to take. 
In section \ref{Sec: Continuum} and \ref{Sec: Deconstructed} 
we show, from the continuum and
deconstructed perspective respectively, that $\Tr   F_{56}^2$ is in fact
generated with appropriate coefficient.  Additionally, in section \ref{Sec: Deconstructed}, 
we discuss the different radiative corrections that arise in the 
deconstructed theory and argue that
we do not lose control of the qualitative dynamics of the theory, even tough we are taking the strong
coupling limit. We also interpret the relation between the continuum
calculation and deconstructed calculation. Conclusions and applications are
then discussed.
 
\section{UV Completing 6 Dimensional Gauge Theory}
\label{Sec: UV Completion}

\begin{figure}[ht]
\centering \epsfig{figure=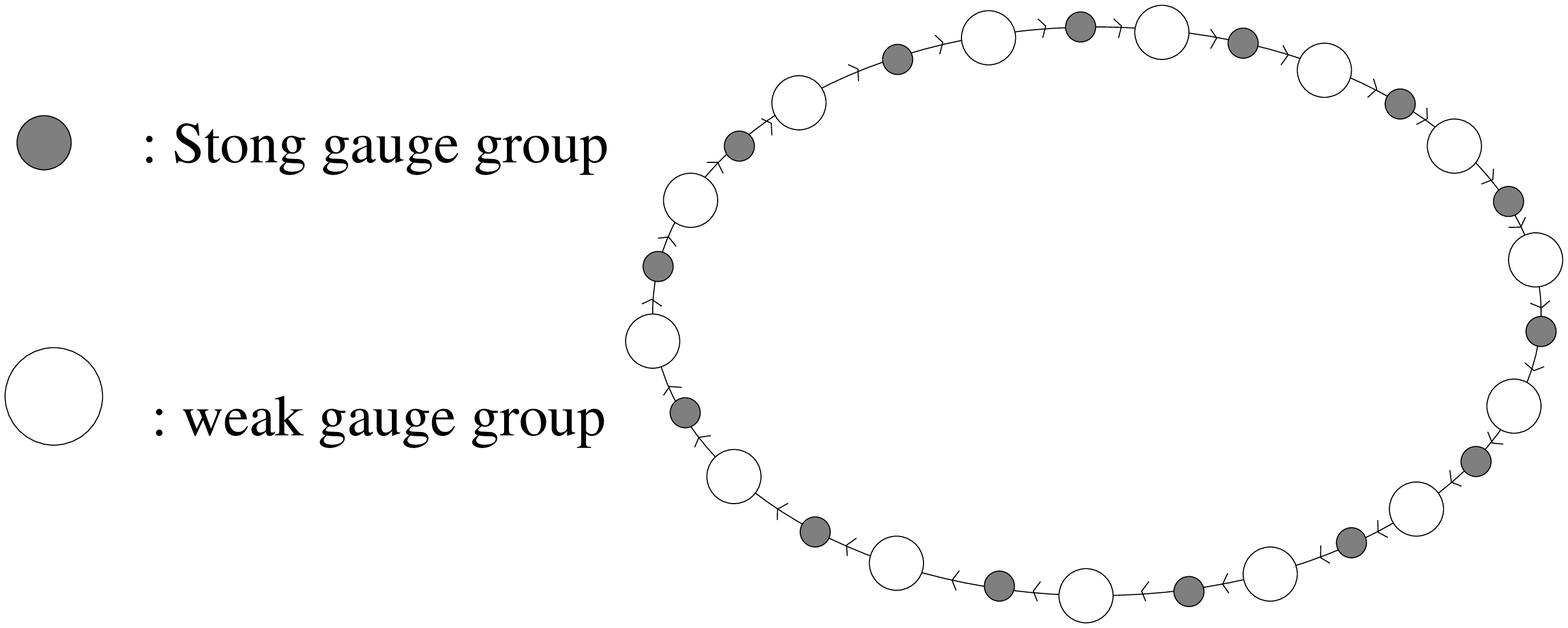, width=7cm}
\hspace{1.0in}
\centering \epsfig{figure=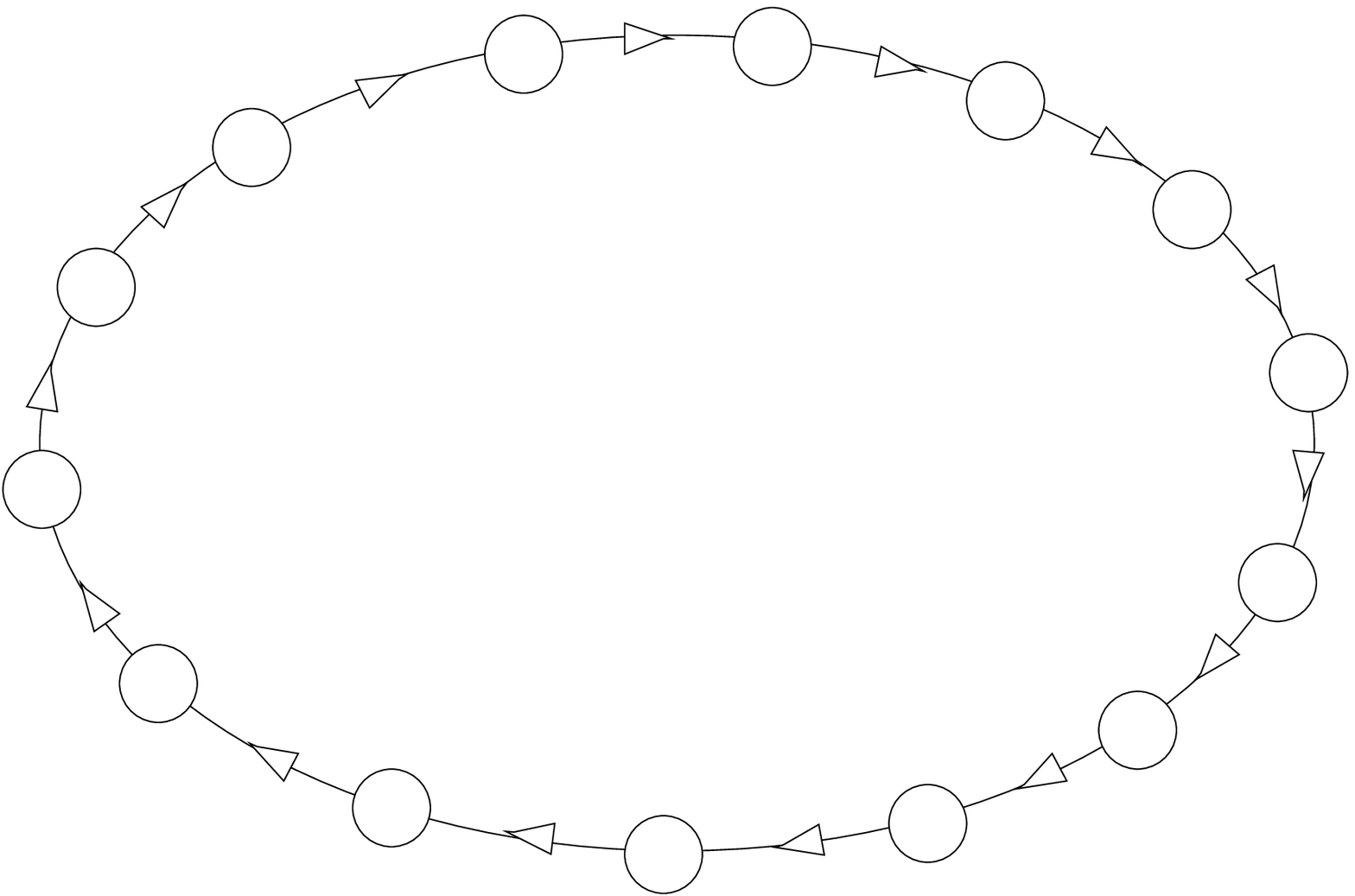, width=5cm}
\caption{Moose diagrams for an ``uncondensed'' and ``condensed''
deconstructed five dimensional theory}
\label{Fig: Circle}
\end{figure}

We would like to review the framework in which extra dimensional gauge
theories can be completed into QCD-like models.
This was first demonstrated for five dimensional theories
in \cite{Arkani-Hamed:2001ca}
and extended to six dimensional theories in \cite{Lane:2002pe}. In the 5D
case, the ultraviolet theory is an $SU(m)^N \times SU(n)^N$ gauge theory,
where the $SU(n)$ gauge groups become strongly coupled at scale
$\Lambda$. The theory has $N$ fermions $\chi_i$ that transform as
$(m,\bar{n})$ under $SU(m)_i \times SU(n)_i$ and N fermions $\psi_i$
transforming as $(n,\bar{m})$ under $SU(n)_i \times SU(m)_{i+1}$. This
structure can be easily visualized in a moose diagram with $2
N$ points and $2 N$ links with the points representing the $SU(m)$ and
$SU(n)$ gauge group alternatively and the links representing the $\chi$'s and the
$\psi$ alternatively (see Fig. \ref{Fig: Circle}). Below the scale $\Lambda$, $\chi_i$ and $\psi_i$
 form condensates. The theory can then be described by a weakly gauged
non-linear sigma model and can be represented by a moose diagram with $N$
$SU(m)$ sites linked by bifundamental non-linear sigma model fields
$U_i$. This non-linear sigma model is a five dimensional gauge theory
compactified on a circle, with the fifth dimension latticized; the moose
diagram has turned into a picture of an extra dimension.  

The naive extension to the 6D case is straightforward. The moose diagram is now a mesh of
alternating weak and strong groups connected by oriented links representing
fermions transforming as fundamental under the gauge group the link is
pointing to and as anti-fundamental under the group the link is pointing
from. Below the scale of strong coupling, the strong gauge groups
``disappear'' from the moose diagram and we are left with $N^2$ weakly
coupled gauge groups $G$ linked by non-linear
sigma model fields $U_{i,j}$ and $V_{i,j}$. These link fields spontaneously break
the $G^{N^2}$ gauge symmetry to the diagonal subgroup, leading to a
collection of massive vector bosons. At low energy, these massive gauge
bosons match part of the ``KK'' spectrum of a six dimensional theory compactified
on a torus.  This is because the kinetic terms for the link fields are the
latticization of the $F_{\mu 5}^2$ and $F_{\mu 6}^2$ terms of a continuous
six dimensional theory. However, the $F_{56}^2$ term corresponding to
plaquette potential is not automatically present. In the rest of this
paper, we show that in an appropriate limit, this term is in fact generated
from infrared physics within this effective non-linear sigma model, with a
coefficient of the appropriate size. Therefore, further discussion of the
ultra-violet completion in term of QCD-like model will not be necessary.   

\begin{figure}[ht]
\centering\epsfig{figure=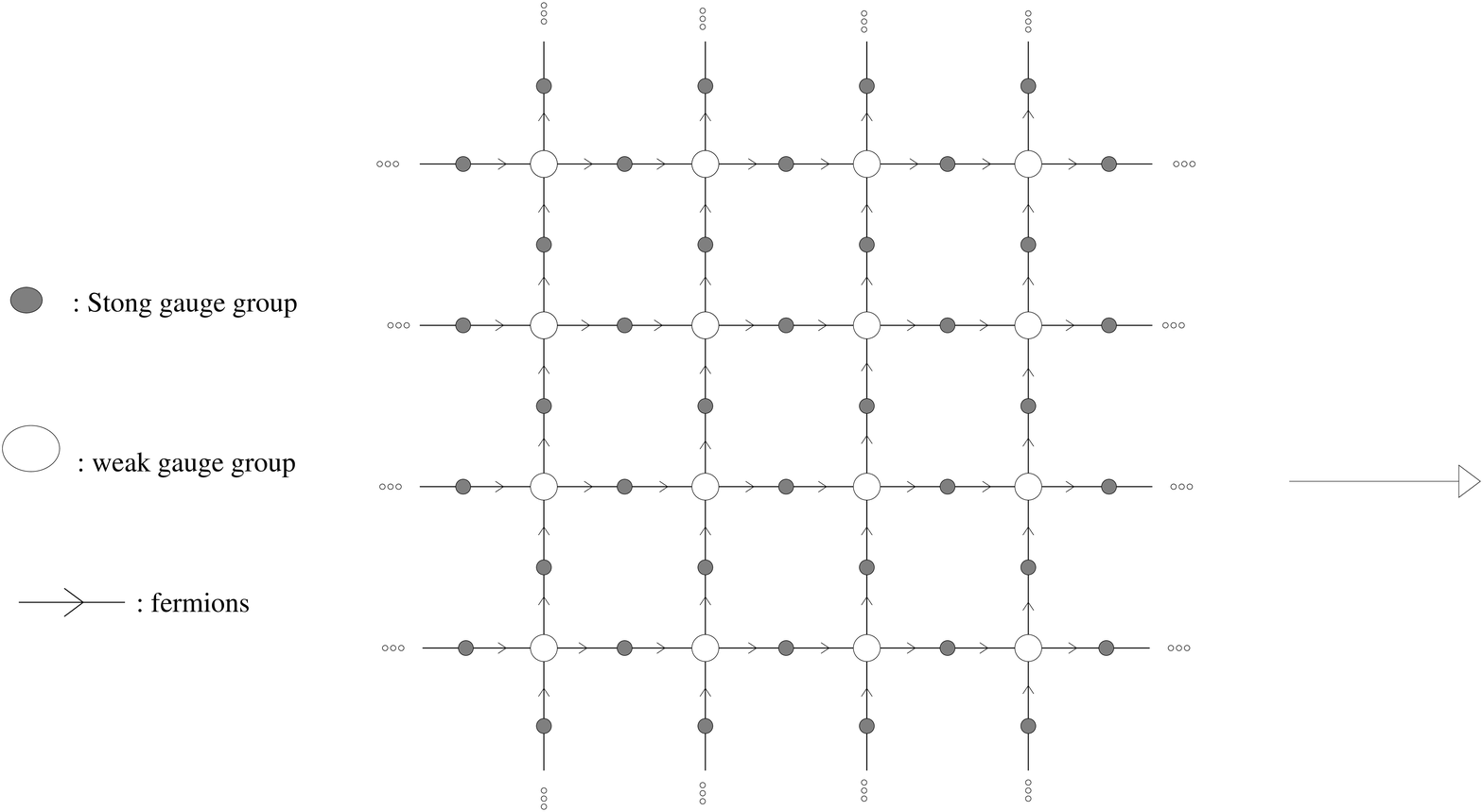, width = 8cm}  
\hspace{0.2in} \centering\epsfig{figure=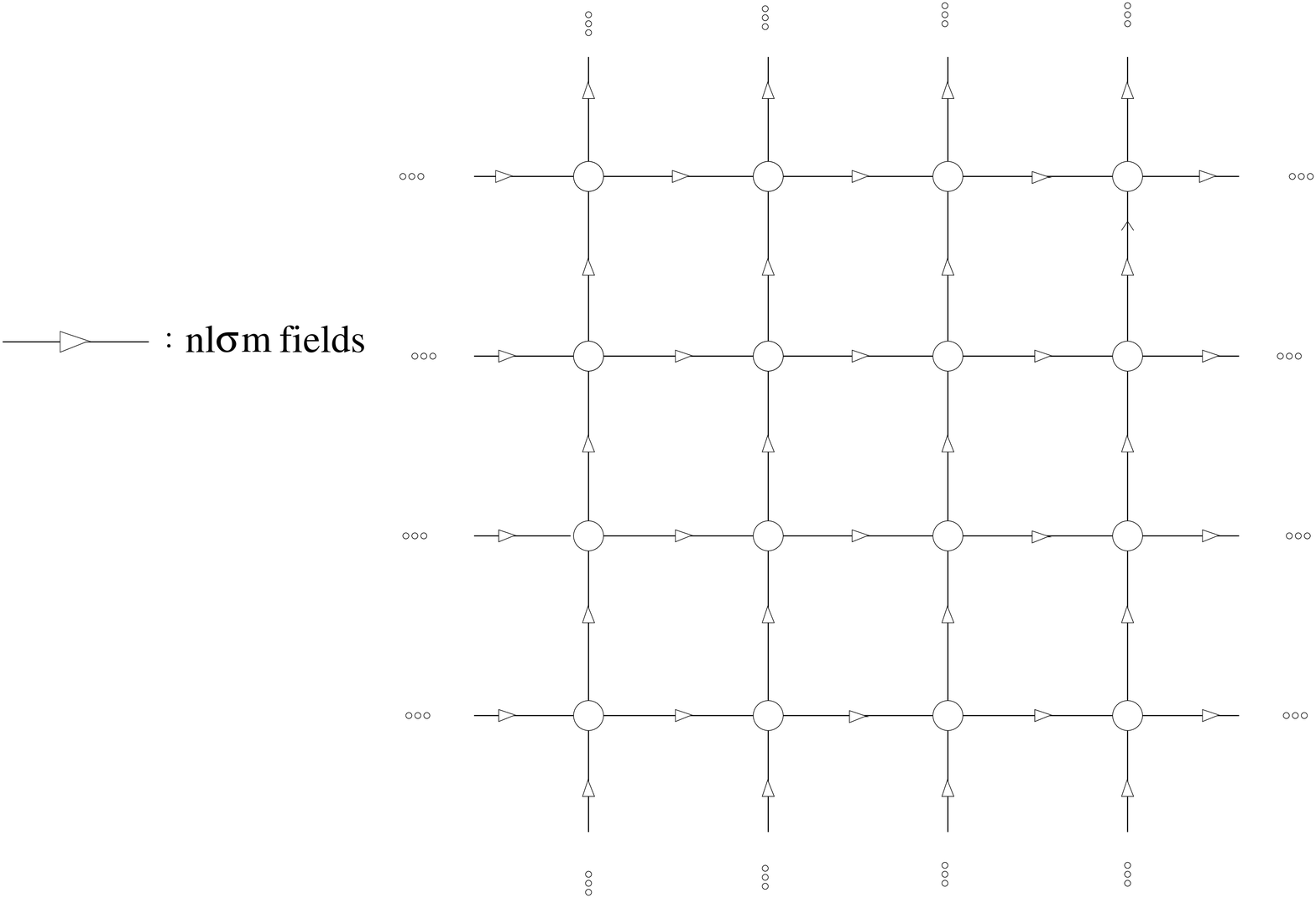, width = 6cm}  
\caption{ Moose diagrams for six dimensional theory
\label{Fig: Mesh}
}
\end{figure}

\begin{figure}[ht]
\centering\epsfig{figure=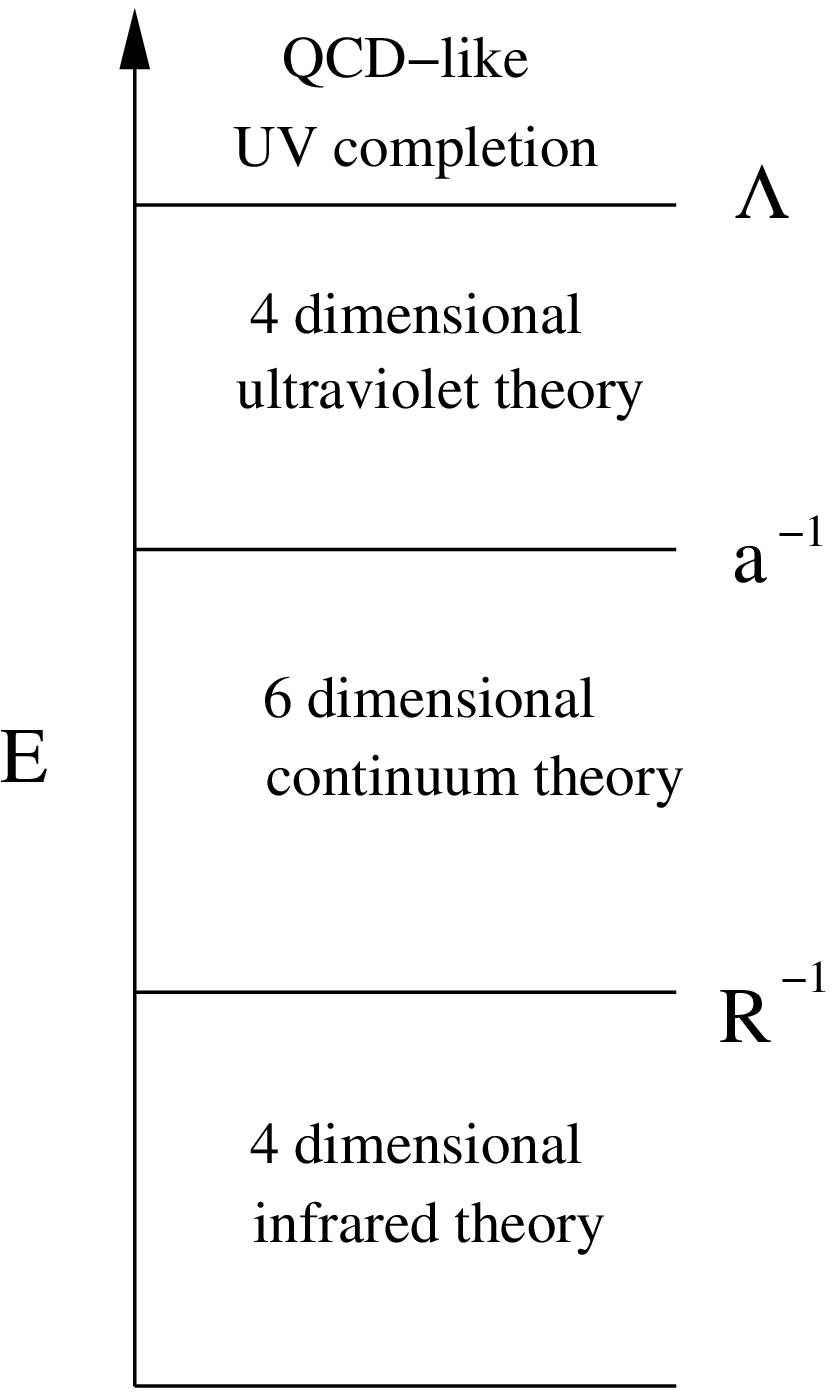, width = 3.5cm}
\caption{
\label{Fig: Scales}
}
\end{figure}

\section{Scaling}
\label{Sec: Scaling}

In this section we discuss the scaling of the couplings
in the different regimes of the theory(see Fig. \ref{Fig: Scales} and establish a dictionary between
the parameters of the four dimensional and six dimensional theories. At the highest
energies, the theory is a four dimensional gauged  non-linear
sigma model with  $N^2$ gauge groups.   The parameters
of this theory are  the breaking scale $f$, the cut-off of
the theory $\Lambda$, the gauge couplings, $g_0=g_0(\Lambda)$, that we
take to be identical, and  the number of sites $N^2$.
The six dimensional continuum description arises at intermediate
energy scales between the lattice spacing, $a$, and the
radius of the extra dimension, $R$.  Between those scales we can define an
effective six dimensional gauge coupling, $g_6$. On it's own, this six
dimensional theory would get strongly coupled at energies of order $4 \pi
g_6^{-1}$, but since it is part of our deconstructed theory, it is regulated
at the scale of the lattice spacing $a^{-1}$.
At energy scales below $R^{-1}$, the theory behaves as
a four dimensional gauge theory with a gauge coupling $g_4$.
The effective four dimensional degrees of freedom below this scale are 
a four dimensional gauge boson and two classically massless
adjoint scalars, $u$ and $v$.

There are relations between the parameters that we state here that can be
derived by matching the KK tower of gauge bosons in a six dimensional theory
with the spectrum of massive gauge bosons of the non-linear sigma model,
and by matching the low energy coupling of both theories.

The cut-off of the four dimensional high energy theory is
roughly $\Lambda \approx 4 \pi f$, set by the scale when
the non-linear sigma model fields become strongly interacting.
The effective six dimensional gauge coupling is related to the high
energy theory as:
\begin{eqnarray}
g_6^{-1} = f
\end{eqnarray}
The lattice spacing is given by:
\begin{eqnarray}
a^{-1} = g_0 f 
\end{eqnarray}
This is the scale where the Kaluza-Klein modes are truncated
and is where we should cut-off the six dimensional continuum
theory.  
The circumference of the extra dimension is given by 
\begin{eqnarray}
R = N a =  \frac{N}{g_0 f} .
\end{eqnarray}
Finally the low energy four dimensional gauge coupling, $g_4$
is related to the other parameters by:
\begin{eqnarray}
g_4 = \frac{g_6}{R} = \frac{g_0}{N}.
\end{eqnarray}

As we vary the parameters in the theory, we want to keep
the low energy parameters fixed.  The two parameters that
we choose to hold fixed are the low energy gauge coupling, $g_4$,
and the radius, $R$.  This means that as we scale $g_0$, we hold
the ratio $g_0/N$ fixed and this imply that $g_6$, $f$ and $\Lambda$
are all fixed as well. We will push the high energy gauge couplings to the limit of 
perturbativity, $g_0 \rightarrow 4\pi$. In this limit, $a^{-1} \sim
\Lambda$, and the theory looks six dimensional up to the cutoff.




\section{Continuum Radiative Corrections}
\label{Sec: Continuum}

\begin{figure}[ht]
\centering\epsfig{figure=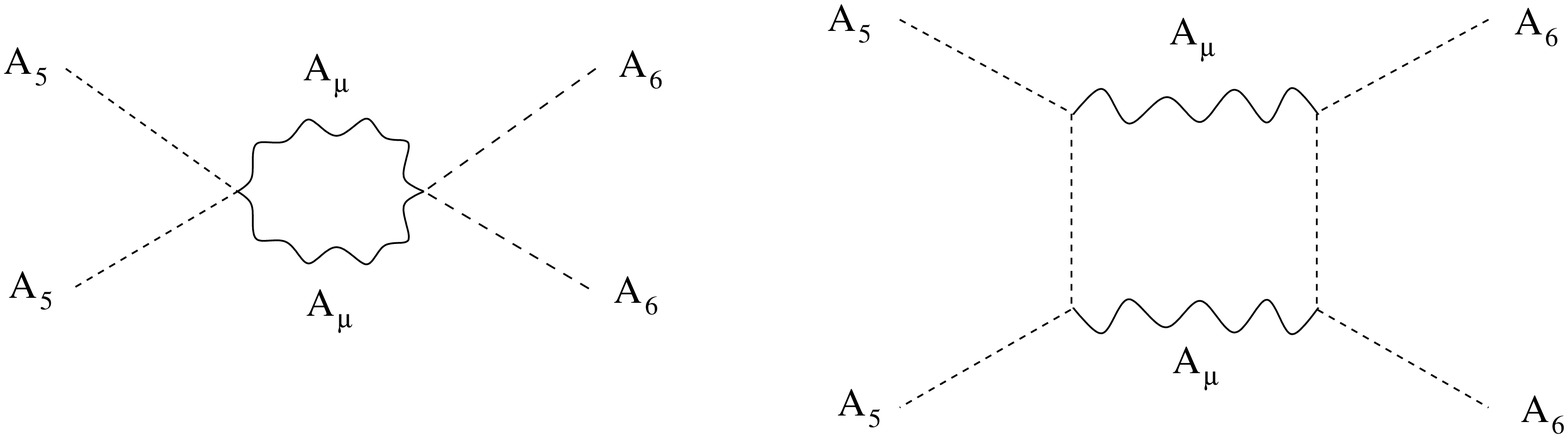, width = 8cm}
\caption{
\label{Fig: 6D diagrams}
}
\end{figure}

The kinetic terms for the non-linear sigma model fields,
$\left|D_{\mu}U_{i.j}\right|^2+\left|D_{\mu}V_{i,j}\right|^2$, corresponds, in the
continuum language, to $\Tr F_{\mu 5}^2$ and $\Tr F_{\mu 6}^2$. The primary issue in
completing the six dimensional theory (or higher dimensional theories as
well), is generating the $\Tr F_{56}^2$ term.  We will generate this term
in the deconstructed theories not from any ultraviolet physics that completes the non-linear
sigma models, but from infrared effects inside
the non-linear sigma model.    We can get a feeling for
how this can happen by considering a six dimensional gauge
theory without Lorentz invariance:
\begin{eqnarray}
S_6 = \int d^6x \Big[ - \frac{1}{2} \Tr F_{\mu \nu}^2
- \Tr F_{\mu 5}^2  - \Tr F_{\mu 6}^2 \Big].
\end{eqnarray}
where the normalization is chosen such that $F_{MN}$ contains the gauge
coupling $g_6$.  We have set the coefficient of $\Tr F_{56}^2$ to zero by hand.
There is no enhanced symmetry when this coefficient vanishes
and therefore it should be generated in the effective action.
The question is what size the coefficient is generated with.
There are two one loop diagrams that generate an $A_5^2 A_6^2$
interaction in the effective action and they are shown in 
Fig. \ref{Fig: 6D diagrams}.  These diagrams are quadratically
divergent in six dimensions and produce a coefficient for
$\Tr [A_5, A_6]^2$ of order: 
\begin{eqnarray}
\sim  \frac{g_6^4}{64 \pi^3} \Lambda^2_6 .
\end{eqnarray}
As mentioned earlier, the six dimensional theory is regulated by having
two of its dimensions on a lattice, we should therefore use the effective lattice
spacing, or more precisely, the top of the heavy bosons spectrum,  as the
cut-off of the six dimensional theory, $\Lambda_6 = 8 g_0^2 f^2$. By gauge invariance, we therefore expect a term:
\begin{eqnarray}
\label{Eq: F56}
 c \frac{g_0^2}{8 \pi^3 } \Tr F_{56}^2.
\end{eqnarray}
to be generated, with $c$ of order $1$. This is precisely the size that we want as $g_0 \rightarrow 4 \pi$.  
It is important to note that we are not claiming that Lorentz invariance
is exactly recovered (as we have left out all
order unity factors), but merely saying that it is unnatural
for the coefficient of $\Tr [A_5, A_6]^2$ to be significantly
smaller than $g_6^2$.  Notice that this term is completely
generated inside the six dimensional theory though it  
is cut-off sensitive.   Dimensional deconstruction provides
an ultraviolet completion for these theories and in the
next section we explore what this ultraviolet completion
says about the quadratic divergence.

$\Tr F_{56}^2$ generates a KK tower as well as quartic couplings for $A_5$
and $A_6$. In the next section, we will check, directly from the
deconstructed theory, that this term is generated with the appropriate
coefficient by looking at the quartic coupling between the massless mode
$u$ and $v$ which correspond to the zero mode of $A_5$ and $A_6$. From
Eq. \ref{Eq: F56}, we expect to find:
\begin{equation}
\label{Eq: F56prediction}
\int d^6x c \frac{g_0^2}{8 \pi^3} g_6^2 \Tr[A_5,A_6]^2 \rightarrow \int d^4
x c \frac{g_0^4}{8 \pi^3 N^2} \Tr [u,v]^2
\end{equation}
Again, as $g_0\rightarrow 4 \pi$, the coefficient of $[u,v]^2$ approaches
$g_4^2$ and become of the same order as the coefficient we would expect
for a Lorentz invariant six dimensional gauge theory. We will also look for
a  tower of massive scalars corresponding to the $KK$
modes of $A_5$ and $A_6$ (denoted $u^{(n,m)},v^{(n,m)}$):
\begin{equation}
m^2_{u^{(n,m)}} = c  \frac{g_0^2}{8 \pi^3} \frac{4 \pi^2}{R^2} \left(n^2 + m^2\right)
\end{equation}

In the continuum theory, we don't expect the zero mode of $A_5$ and $A_6$
to remain exactly massless. At low energy, the theory looks four
dimensional, and contains two massless scalars $u$ and $v$, as well as a
massless gauge boson. This massless gauge boson will generate a potential
for $u$ and $v$  that is not gauge invariant in an infinite six
dimensional theory. This potential  comes from infrared physics and is therefore finite
and calculable\cite{Antoniadis:2001cv} . We won't calculate it exactly, but we can estimate it from the physical picture: It is the
effective potential of the low energy theory, cutoff at the scale of the
first KK mode $m_{KK}$. This effective potential can be  obtained by turning on a
background for $u$ and $v$, calculating the mass matrix of the gauge bosons
$M(u,v)$ in this background and using the Coleman-Weinberg formula with
$m_{KK}$ as the cutoff:
\begin{equation}
V(u,v) = \frac{3}{64 \pi^2}\left( \Tr M^2(u,v) m_{KK}^2 + \Tr M^4(u,v) \log \frac{M^2(u,v)}{m_{KK}} \right)
\end{equation}
For $u$ and $v$ turned on in the same direction ($\sigma_z$ for example), we get:
\begin{eqnarray}
\label{Eq: Cwlowenergy}
&&\frac{6}{64 \pi^2}\left(  \frac{4 g_4^2 \pi^2}{R^2} (u^2+v^2) + g_4^4
(u^2+v^2)^2 \log\left(\frac{ g_4^2 \left(u^2 +v^2\right) R^2}{4 \pi^2} \right)\right)\\
\nonumber
&=& \frac{6}{64 \pi^2}\left( \frac{4  g_4^2 g_0^2  \pi^2}{N^2} (u^2+v^2) f^2 + g_4^4
(u^2+v^2)^2 \log\left( \frac{ g_4^2 N^2 \left(u^2 +v^2\right) }{g_0^2 f^2 4 \pi^2} \right)\right)
\end{eqnarray}
We will see in the next section, that this result is reproduced closely by
the deconstructed calculation when $N$ is large. This part of the
discussion is very similar to the five dimensional case \cite{Arkani-Hamed:2001ca}.

\section{Deconstructed Calculation}
\label{Sec: Deconstructed}

Having seen how the six dimensional continuum theory
naturally generates a large quartic potential,
we wish to find out how this arises in the deconstructed
completion.  In doing so we will find that the quadratically
divergent coefficient will actually arise from a finite
effect in the deconstructed picture.  

\begin{figure}
\centering \epsfig{figure=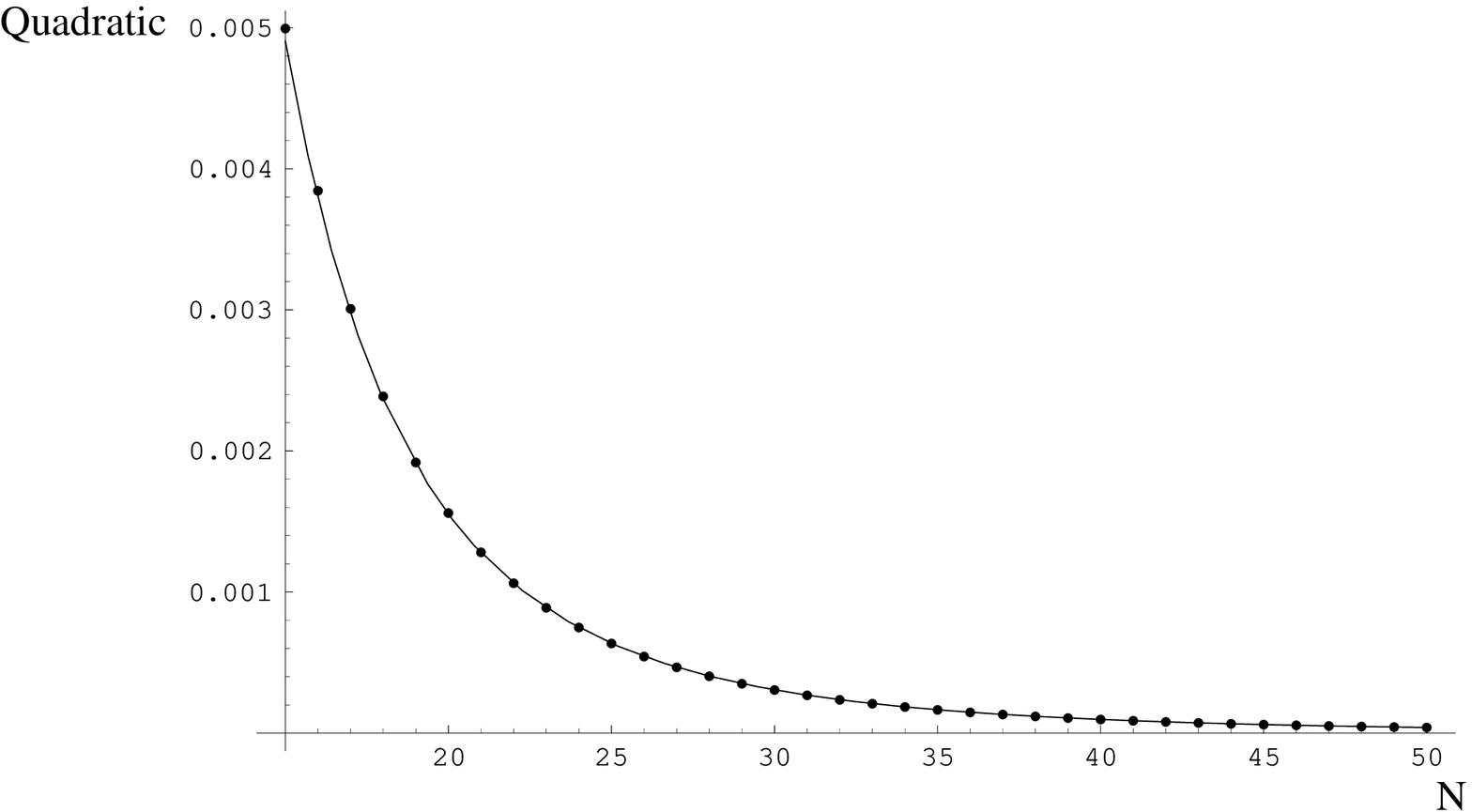,width=7cm} \hspace{1cm} \epsfig{figure=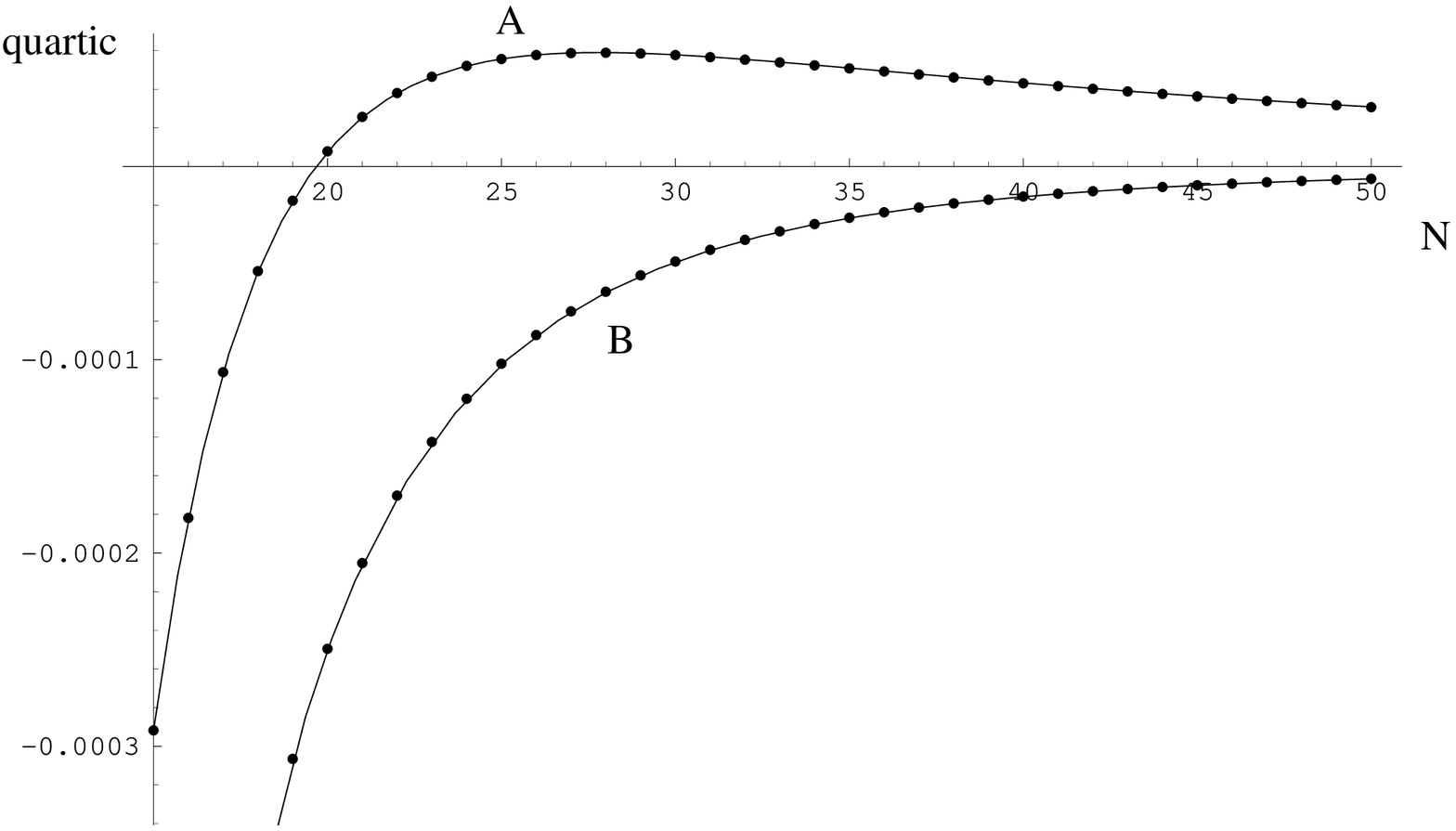, width=7cm}

\caption{Numerical calculation of the Coleman-Weinberg potential, keeping
$g_0$ fixed. The Plot
to the left, represent the mass as a function of $N$, fitted to a
$X/N^4$ curve, where $X$ determined from the points.
The set of 
points A on the right hand side plot shows the quartic coupling ($\lambda + \kappa_1$) as a function of $N$ when the zero mode
backgrounds are turned on in two different directions. They are fitted to a $Y/N^2 + Z/N^4$ curve, where
$Y$ and $Z$ are determined from the points. The lower set of points (B), is the
quartic coupling  ($\kappa_1+\kappa_2$) when both zero mode are turned on in the same
direction. They are fitted to a curve of the form $W/N^4$. }
\label{Fig: Plots}
\end{figure}

We will consider our theory beneath the scale of condensation
where the non-linear sigma model is the appropriate description
of physics.  The Lagrangian for this theory is given by:
\begin{eqnarray}
\LL_{\text{nl$\sigma$m}} 
= \sum_{a,b}^{N} \Big[
- \frac{1}{2} \Tr F^2_{(a,b)}  
+ \frac{f}{4} \Tr \big| D_\mu U_{(a,b)}\big|^2
+ \frac{f}{4} \Tr \big| D_\mu V_{(a,b)}\big|^2\Big]  + \cdots.
\end{eqnarray}
where the ellipses represent higher derivative operators and
heavy cut-off scale states.  For calculational simplicity, we will
consider an $SU(2)$ gauge theory, however the general arguments
will apply to any non-Abelian gauge group. We want to demonstrate that
a significant quartic potential is generated for the 
constant modes of the theory:
\begin{eqnarray*}
U_{(a,b)} = \exp\big(i u / Nf\big)
\hspace{0.7in}
V_{(a,b)} = \exp\big(i v / Nf\big) .
\end{eqnarray*}
At one loop there is no quadratic divergence or logarithmic
divergence for these modes
because the breaking of the chiral symmetries that protect
these pseudo-Goldstone bosons necessitate many gauge couplings and
ultraviolet physics is analytic in the couplings (see \cite{Arkani-Hamed:2001nc}).  However,
there is an infrared contribution to both the mass of these
modes as well as the quartic coupling. These contributions can be calculated
from the 1-loop Coleman-Weinberg potential . We first turn on background
fields proportional to $\sigma_z$ for both $u$ and $v$. The masses squared
of the gauge bosons in that background are given by: 
\begin{equation}
M^2_{n,m}(u,v) =4 g_0^2
f^2\left( \sin^2 \frac{2 n \pi + u}{2 N}+ \sin^2 \frac{2 m \pi + v}{2 N}\right)
\end{equation}
 The Coleman-Weinberg potential is given by:
\begin{equation}
V(u,v) = \frac{1}{64 \pi^2}\sum_{n,m} M^4_{m,n}(u,v) \log \frac{M^2_{m,n}}{\Lambda}
\end{equation}
The mass and quartic coupling can be extracted from the potential above:
\begin{eqnarray}
m^2_{u} = \frac{\partial^2}{\partial u^2} \sum_{n,m}  M^4_{n,m}(u,v)
\log\frac{M^2_{n,m}(u,v)}{\Lambda}\Big |_{u,v=0} \\
\nonumber
\kappa =\frac{1}{4!} \frac{\partial^4}{\partial u^4} \sum_{n,m}  M^4_{n,m}(u,v)
\log\frac{M^2_{n,m}(u,v)}{\Lambda}\Big |_{u,v=\phi} 
\end{eqnarray}    
Note that for the quartic term, we have given $u$ and $v$ a vev $\phi$ in
order to avoid infrared divergence. We computed  the masses and
the quartic couplings numerically from these
expressions. We then repeated the procedure for $u$
and $v$ proportional to $\sigma_x$ and $\sigma_y$ respectively. The masses
of the gauge bosons in this background are more complicated, but we can still
compute the quartic couplings numerically. The quartic couplings in this
case contain a contribution from $\Tr[u,v]^2$ which
doesn't contribute when both $u$ and $v$ are turned on in the same
direction. The results of the numerical calculation are shown in
 Fig. \ref{Fig: Plots} and can be expressed as: 
\begin{eqnarray}
V_{\text{1 loop}} 
= m^2 \Tr\big( u^2 + v^2\big) 
+ \lambda \Tr [ u, v]^2  + \kappa_1 \left(\Tr u^4+\Tr v^4 \right)
+ \kappa_2 \left(\Tr u v\right)^2 
\end{eqnarray}
with: 
\begin{eqnarray}
m^2 \sim  \frac{g_0^4 f^2}{16 \pi^2 N^4}
\hspace{0.5in}
\kappa_{1,2} \sim \frac{g_0^4}{16 \pi^2 N^4} 
\hspace{0.5in}
\lambda \sim  \frac{g_0^4}{16 \pi^2 N^2} = g_4^2 \frac{g_0^2}{16 \pi^2} .
\end{eqnarray}
We find that our numerical calculations for $m$ and $\kappa$ agree well
with the expectation from the low energy effective potential of the continuum
theory, Eq. \ref{Eq: Cwlowenergy}. Also, our result for $\lambda$ agrees
with the divergent contribution of the continuum theory, Eq. \ref{Eq: F56prediction}. These results should be viewed as parametric relations only, as many
effects can modify the order $1$ coefficients involved in these
expressions. For example, the  operators generated from UV physics that will
be discuss next can have
important effects, and the running of the couplings can change the naive relation
between the high energy coupling $g_0(\Lambda)$ and $g_4$.

\begin{figure}
\centering \epsfig{figure=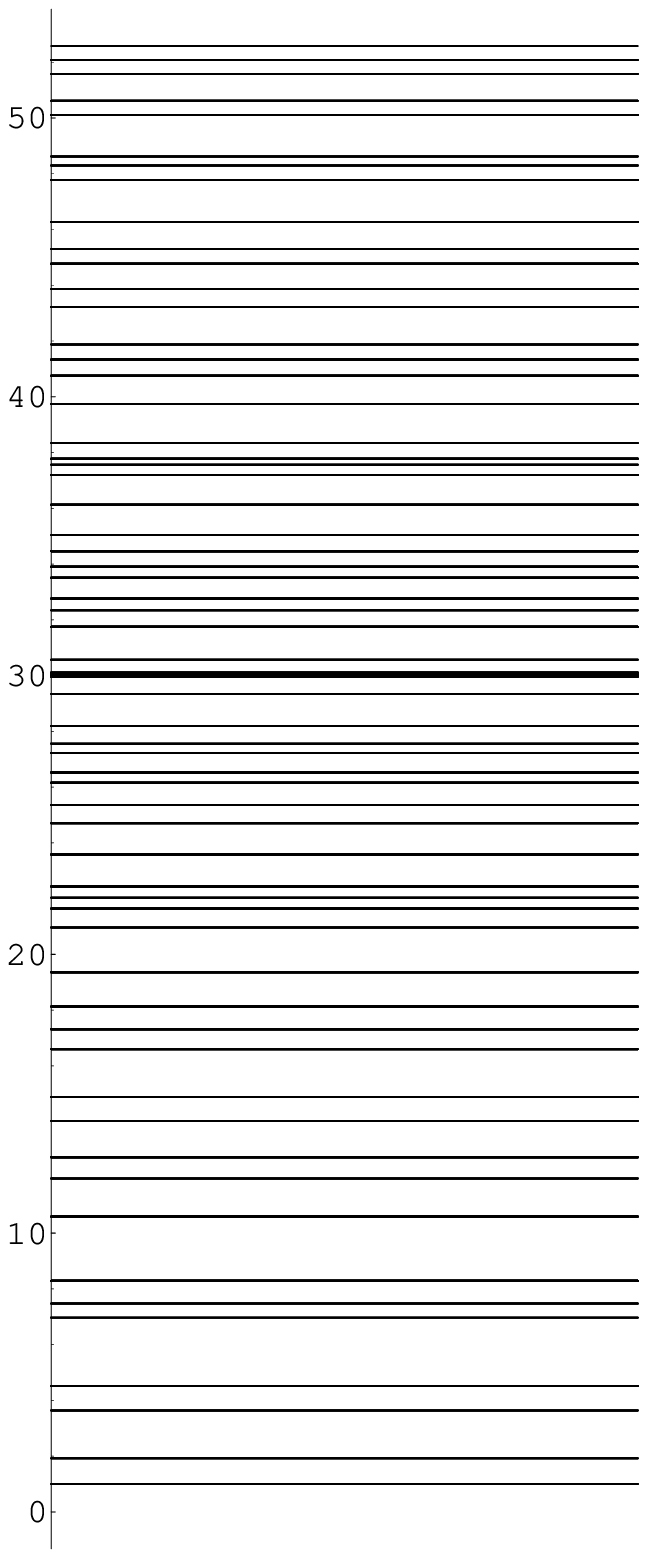, width=3cm} \hspace{0.3in}
\epsfig{figure=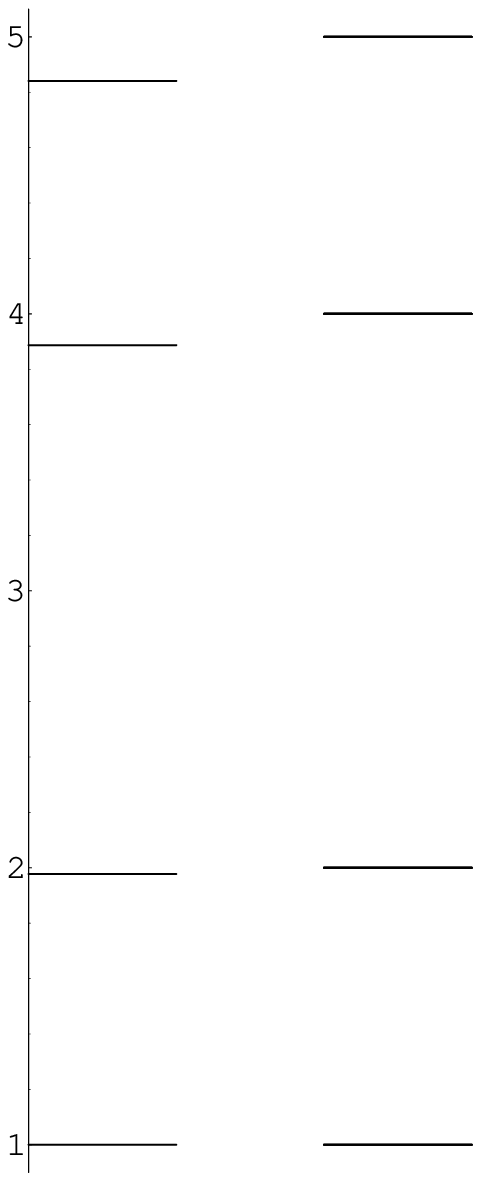, width = 3cm}
\caption{The plot on the left shows the masses of the (non zero mode)scalars, computed numerically for $N=20$. The right-hand side plot
shows a comparison between the first 4 massive modes of the deconstructed
theory for $N=40$ (on the left) and a continuum KK spectrum (on the right)}
\label{Fig: KK}
\end{figure}

On the continuum side, the $\Tr F_{56}^2$ term also
contains $(\partial_5 A_6 - \partial_6 A_5)^2$ which should correspond
on the deconstructed side to a tower of massive scalars which should also
appear in the finite one loop Coleman-Weinberg potential. In Fig. \ref{Fig: KK},
we show the masses of the scalars of the theory, generated by the one-loop
Coleman-Weinberg potential which was computed numerically for $N=20$. We
find that a tower of massive modes is indeed generated. We also show the
first $4$ massive modes for $N=40$ compared to the KK spectrum of a
continuum theory. They are in reasonable agreement, given the relatively
low value of $N$.  

Until now we have talked only about radiative corrections coming from
infrared physics. There are also UV contributions that are not calculable,
but can be estimated from spurion analysis and naive dimensional
analysis. We now discuss those contributions that could potentially become
large as we take the limit of strong coupling. Plaquette operators will be
generated with quadratically divergent coefficient at four loops. Using \mbox{ $\mathcal{P}_{ij} =
U_{(i,j)} V_{(i+1,j)} U_{(i+1,j+1)}^\dagger V_{(i,j+1)}^\dagger$}
as the $(i,j)$th plaquette, the effective action contains a
term:
\begin{eqnarray*}
\delta \lambda_{ij} |\Tr \mathcal{P}_{ij}|^2 \sim 
16 \pi^2 f^4 \left( \frac{g_0^2}{16 \pi^2}\right)^4
| \Tr \mathcal{P}_{ij} |^2
\end{eqnarray*}
This results in an contribution to the quartic coupling of:
\begin{eqnarray*}
\label{Eq: quarticUV}
\delta \lambda \sim \frac{16 \pi^2 }{N^2} \left(\frac{g_0^2}{16
\pi^2}\right)^4= g_4^2 \left(\frac{g_0^2}{16 \pi^2}\right)^3
\end{eqnarray*}
where we have taken into account a factor of 
$(N f)^{-4}$ for the normalization of the zero modes and the fact that
there are $N^2$ plaquettes.  
This quadratic divergence has the same $N$ scaling
as the infrared contribution, and is roughly of
the same size as $g_0\rightarrow 4 \pi$, though it is
suppressed for $g_0 \lsim 4\pi$.
Therefore we can not calculate the exact coefficient of the
quartic coupling, only say that it is not natural for it
to be small. 

There are also quadratically divergent $N$ loops contributions to the zero
mode mass from ``Wilson loop operators'': 
\begin{eqnarray}
\label{Eq: wilsonloop}
\Lambda^2 f^2 \left(\frac{g_0^2}{16 \pi^2}\right)^N \left|\Tr U_{(1,1)} U_{(2,1)}
\cdots U_{(N,1)} \right|^2
\end{eqnarray}
 From the extra dimensional point of view, because they are non
local, these operators should be exponentially small in the size of the extra
dimension. In the deconstructed picture, for $g_0 \lsim 4 \pi$ they are also
exponentially suppressed. For any value of $g_0$ close to $4 \pi$, we can
increase $N$ so that the contribution of Eq. \ref{Eq: wilsonloop}
is arbitrarily small while the quartic couplings  \mbox{Eq. \ref{Eq: quarticUV}} and
Eq.\ref{Eq: F56prediction}, take the appropriate value for a six
dimensional theory. More generally, if we take $g_0 = 4\pi/(1+\epsilon)$,
operators that stretch over $n$ sites (``large plaquettes''), and correspond
to non-local operators will have a size
$\sim e^{-4 n \epsilon}$, while the local operators Eq. \ref{Eq: quarticUV}
 and Eq.\ref{Eq: F56prediction} will have size $g_4^2
(1-4 \epsilon)$ and $g_4^2 (1-2 \epsilon)$, which is what is expected for
local operators of a six dimensional theory. This shows that,
parametrically, we generate with  appropriate coefficients the
operators that we expect from a local six dimensional theory with departure
from Lorentz invariance of order $\epsilon$, while the
operators that are non-local over distances larger than $\sim a/\epsilon$
are exponentially suppressed. Therefore, for $g_0$ near $4\pi$, we generate
a theory that parametrically looks like a six dimensional gauge theory with
departure from Lorentz invariance of order $\epsilon$, that is local at
distance scales larger than $\sim a/\epsilon$ and smaller than $R$. We can then take $N$
large, increasing the radius, so that there is always a range in which the theory looks like a local six
dimensional theory.

The question of whether or not this can be done for realistic theory, where
$g_4$ is not too small (and therefore $N$ not too big) cannot really be addressed. As we already mentioned
earlier, the size we derived for the various radiative corrections are to
be viewed as parametric relations only. Order one coefficients for the
operators generated in the UV cannot be predicted while even the
computation of infrared effect can be modified, as the naive relation
between $g_0(\Lambda)$ and $g_4$ is affected by ``power law'' running of $g_0$.

\subsection{Comparison with Continuum Calculation}

We have shown that \begin{it}infrared\end{it} effects in the
deconstructed theory generate the quartic potentials in
the large $N$ limit with a suitably large coefficient.
In the previous section we showed that the quartic coupling
arose as a quadratically divergent contribution in
the continuum limit.  One might worry that this is some sort
of contradiction.   In the limit we are taking, $g_0 \rightarrow 4\pi$,
the high energy regime where the theory looks four dimensional
shrinks and disappears.  Referring to Fig. \ref{Fig: Scales},
we see that in this limit the theory looks six dimensional up to the cutoff
at the scale of strong coupling $\Lambda$.  The infrared contributions to the
quartic couplings arise from the appearance of ``Kaluza-Klein''
bosons in the deconstructed theory.  The quadratic divergence
in the continuum theory also arise because of the increasing
number of Kaluza-Klein gauge bosons.   Therefore, the contributions
to the quartic potential have the same origin in both the deconstructed
picture and the continuum picture.

\section{Conclusion}

In this paper, we have demonstrated that it is possible to dynamically
generate six dimensional gauge theories from four dimensional
asymptotically free gauge theory. This generalize the construction of 
\cite{Arkani-Hamed:2001ca}.
The new element in deconstructing six
dimensional (as well as higher dimensional) theories is the presence of the $\Tr
F_{56}^2$ term which generate a KK spectrum for $A_5$ and $A_6$ and
contains quartic interactions between higher dimensional
components of the gauge field, which translate in the deconstructed theory
into quartic interactions between the pseudo-Goldstone bosons. This term
was shown to be
generated naturally with appropriate coefficient in deconstructed theories
when we take the limit of strong coupling.  It
arise not from operators in the ultraviolet completion
of the theories, but through infrared effects generated
in the regime where the theory looks extra dimensional.
This allows us to consider QCD-like theories
as ultraviolet completions for six dimensional theories,
as well as higher dimensional theories. We note however that the continuum
theory that emerges  is
not exactly Lorentz invariant as we have no control on the exact
coefficient with which $\Tr F_{56}^2$ is generated.

We have shown this result from two different but equivalent
perspectives. From the point of view of the 6 dimensional continuum theory,
$\Tr F_{56}^2$ is generated with quadratic divergences that are regulated by
the deconstructed theory at the scale of the lattice spacing. As we take
the high energy gauge coupling to be large, the coefficient of
this term becomes of the right order. From the perspective of the
deconstructed theory, $\Tr F_{56}^2$ contains a $[u,v]^2$ term which is generated by finite one
loop effect and is seen from numerical calculation to agree with the
continuum estimate. We have also seen that a tower of massive scalars is
generated by finite one loop effect, corresponding to the KK tower of $A_5$
and $A_6$. 

Finally, we have shown that operators of the deconstructed
theory that stretch in theory space and that correspond to non-local
operators in the continuum description, can be kept small, even as we
approach strong coupling, by making $N$ large. How large exactly it can be
is however a question that we cannot answer since it would require precise
matching to the strongly coupled theory.

One application of this result, is the possibility of having
QCD-like UV completion of little Higgs model based on theory
space \cite{Arkani-Hamed:2001nc,Arkani-Hamed:2002pa}. In these theories,
the Higgs are the zero mode $u,v$  and they get
their quartic coupling from plaquette potential. The success of these models
rely on the fact that the Higgs get a small mass from radiative corrections
and  large quartic couplings from the plaquette that are put in at tree
level. We can build realistic
model where the Higgs is a fundamental of $SU(2)$ by using a moose made out of
$SU(3)$ sites except one which is taken to be $SU(2) \times U(1)$. The low
energy gauge group is then $SU(2) \times U(1)$, and the zero mode scalar
that were in adjoint of $SU(3)$ decompose into doublets, triplets and singlets
of $SU(2)$. 

Generating the plaquette potential
was seen as an obstacle to completing the little Higgs theories with
strongly coupled dynamics. But we have shown in this paper that in the
appropriate limit $g_0 \rightarrow 4 \pi$, a quartic coupling of order
$\lambda \sim 16
\pi^2/N^2 \sim g_4^2$ is generated while the mass of the Higgs is given
parametrically by $m_{\text{Higgs}}^2 \sim \left(g_4^2/16 \pi^2\right)^2 \Lambda^2$.  Note, that the quartic
coupling for the Higgs is of order $g_4^2$ much as in the MSSM and it can be
kept small as we make $g_0$ large by taking the number of sites $N$ large. We
should note  that there are other  ways of
generating quartic couplings for little Higgs, either through Yukawa
interactions in theory space models \cite{Gregoire:2002ra,Arkani-Hamed:2002qx} or
through gauge dynamics in coset models \cite{Arkani-Hamed:2002qy}. We
haven't discuss how Yukawa couplings could be generated from strong
dynamics. At low energy, introducing Yukawa interactions that do no spoil
the ultra-violet insensitivity of the Higgs mass (see\cite{Arkani-Hamed:2001nc,Arkani-Hamed:2002pa,Arkani-Hamed:2002qx} ) requires the generation of
operators coupling link fields and fermions. They could for example arise
from four fermion operators as in ETC models. This point offers interesting
avenues of investigations.

\section*{Acknowledgments}
We wish to thank Nima Arkani-Hamed for very valuable
insights. This work is supported in part by the Department of Energy
under Contracts DE-AC03-76SF00098 and the
National Science Foundation under grant PHY-95-14797. T. Gregoire is also
supported by an NSERC fellowship.
\providecommand{\href}[2]{#2}\begingroup\raggedright

\endgroup
\end{document}